\documentclass[oneside,a4paper]{article}

\usepackage{hyperref}
\usepackage{float} 
\usepackage{tikz}
\usepackage{pgfplots}
\pgfplotsset{width=10cm, compat=1.9}

\usepackage{authblk}
\providecommand{\keywords}[1]{\textbf{\textit{Keywords:}} #1}

\newtheorem{problem}{Problem}
\newtheorem{example}{Example}
\newtheorem{definition}{Definition}

\title{An outline of multi objective optimization in databases with focus on flexible skyline queries}
\author{Matteo Savino}
\affil{Politecnico di Milano\\
Milan, Italy\\
\href{mailto:matteo.savino@mail.polimi.it}{matteo.savino@mail.polimi.it} }
\date{January 13, 2022}

\begin{document}
\maketitle
\begin{abstract}
The problem of optimizing across different, conceivably conflicting, criteria is called multi-objective optimization and it is widely spread across many fields. This is a recurring problem in database queries when there is the need of obtaining the best objects from a very large data set. In this article, I included a complete review of the main approaches typically used to achieve multi-criteria optimization. Starting from ranking queries and skylines and then proceeding to more advanced methods, this paper aims to define a clear outline of multi-objective optimization in databases. In particular, the flexible skyline paradigm is considered and thoroughly discussed as it overcomes many of the critical issues that arise with other methods.
\end{abstract}


\section{Introduction}
The problem of optimizing across different, conceivably conflicting, criteria is called multi-objective optimization. As reported in \cite{article8} the need to find the optimal solution to many desired goals is found across many research fields such as mathematics, economics, engineering and many others. Several reviews on this topic have been conducted since it is widespread across multiple fields and many techniques have been designed to solve this problem. Some of these methods do not require complex mathematical equations, therefore they can be easily applied. These simpler approaches fall under two main categories: pareto and scalarization \cite{article8}.

One of the fields in which multi-objective optimization appears very frequently is database queries. \cite{article15} highlights how the modern web infrastructure, together with social networks, produces an unprecedented mass of data. Therefore, data is usually available in very large quantities and hence determining the most relevant items in data sets is fundamental. A slightly more formal definition of the problem has been reported below.
\begin{problem}
Given N objects described by d (dimensions) attributes in a data set and given some notion of quality, possibly conflicting, for each attribute, find the best objects.
\end{problem}
Problem 1 will be dealt with in this survey by showing some of the most widely recognized or innovative techniques. Multi-objective optimization in data sets is commonly obtained with three main approaches \cite{article1, article3, article6}:
\begin{enumerate}
  \item The ranking or top-k (or scalarization as previously mentioned) approach: it reduces the multi-objective problem to a single objective one by using a weighted formula. The weights express the relative importance of each attribute.
  \item The lexicographic approach: It enforces a strict priority among attributes.
  \item The skyline (or Pareto as previously mentioned) approach: It is based on the concept of dominance and it returns all objects that are not dominated by other objects.
\end{enumerate}
The lexicographic method is not covered by this document in detail. However, its disadvantages are clear, a tuple may be better than another across all attributes but the one with the highest priority and be discarded.

In this paper, an overview of the ranking and skyline methods is provided along with a description of their pros and cons. Then, to overcome some of the critical issues, a brief explanation of some evolved versions is given. Finally, the main and final goal is to go into details of one of the previously mentioned evolutions which is the flexible skylines paradigm. In short, it consists of a mix between the top-ks and skylines which aims to get the best of the two worlds.

The structure of the paper is organized as follows. Section 2 discusses ranking queries while section 3 focuses on skylines. Section 4 contains the list of some newer approaches and finally section 5 presents an entire module devoted to flexible skylines. The aim of all sections before the fifth one is to provide the reasons that made this approach necessary. F-skylines are discussed in detail since they are the core topic of the document. Finally, section 6 concludes the paper.

\section{Ranking queries}
Ranking queries, also called top-k queries, have been studied extensively and are by far the most used approach for multi-criteria decision making. The definition of Problem 1 has to be adapted to properly reflect what top-k queries try to achieve.
\begin{problem}
Given N objects described by d (dimension) attributes in a data set and given some notion of quality, possibly conflicting, for each attribute, find the best k objects ordered according with their relevance.
\end{problem}
In Problem 2, the notion of ranking is implicitly introduced and it represents the arranging of objects accordingly with their significance. The importance of a tuple with respect to another one, concept that defines the ranking, is obtained by applying a so called scoring function.

The basic idea behind ranking queries is solving the multi-optimization problem by transforming it into a single optimization one through a weighted formula \cite{article6}, called scoring function \cite{article3}. The most common and widespread cases involve linear scoring functions, hence they will be the focus of the discussion. The definition of the formula is as follow:
\begin{equation}
F=w_{1}f_{1}+w_{2}f_{2}+..+w_{n}f_{n}
\end{equation}
where $F$ is the candidate' score, $w_{i}$ is the weight assigned to the $i$-th criteria $f$ and $n$ is the number of evaluation criteria \cite{article6}. It's clear that weights are an input that need to be either explicitly specified or computed somehow. There are different know methods to obtain weights as shown in \cite{article11}:
\begin{enumerate}
  \item The user is asked to insert directly the parameters or more conceivably some preferences from which weights can be easily obtained. 
  \item User's activities are monitored and weights are learned from their habits.
  \item The preferences of multiple users are aggregated to obtain the needed parameters (e.g. using crowdsourcing \cite{article15}).
\end{enumerate}
An intuitive example of the top-k approach is provided without using a specific algorithm; it will be useful also for pointing out some details further in this paper.

\begin{table}[H]
\centering
\begin{tabular}{||c c c||} 
 \hline
 Student & Math grade & Eng grade  \\ [0.5ex] 
 \hline\hline
 Alex & 9 & 5 \\ 
 James & 10 & 1 \\
 Emma & 6 & 6 \\
 Robert & 6 & 4 \\
 John & 4 & 9 \\ [1ex]
 \hline
\end{tabular}
\caption{Simple data set for examples.}
\label{table:1}
\end{table}

\begin{example}
Consider the data set of Table 1 and the scoring function $StudentScore = 0.5*MathGrade + 0.5*EngGrade$. We want to compute the top-2 query. The computation procedure is shown only for one of the students, the rest can be easily obtained in the same way: $AlexScore = 0.5*9 + 0.5*5 = 7$. The ranking for $k=2$ clearly is: Alex in first place and John in second place.
\end{example}

\subsection{Pros and cons}
Top-k queries are clearly a very simple, efficient and easy-to-use approach and this probably explains why it is widespread. In addition, the ability to control the size of the output and the trade-offs among attributes should not be underestimated \cite{article1, article2, article3, article12}. However, this method is not free of drawbacks and can have substantial disadvantages. First, ranking queries are essentially aggregating different criteria and thus they are possibly mixing up different units of measurements or, which is worse, non-commensurable criteria \cite{article6}. Data normalization is certainly a possible solution but it requires specific background knowledge and hence an ad-hoc solution for every domain. This inconvenience may prevent its adoption in many areas that require general solutions.

Last but not least, the elephant in the room is the setting of weights. First of all, it is proven that it's difficult to obtain the right values for multiple reasons such as the fact that the user who inputs the data may not themselves know their preferences. Moreover, it is really hard to estimate what change could cause the modification of the weight's vector \cite{article3}. The only way to do it is to experiment by running it several times with different values for the parameters, which evidently is not convenient and once again would end up in an ad-hoc approach. Due to these facts, both making the user inputs the parameters directly and computing them from user preferences does not lead to a perfect solution. As shown in \cite{article15}, some crowdsourcing techniques proved helpful for overcoming this weakness at least to some extent. Unfortunately, this is not enough to consider the problem solved because once a combination of weights is found, then it's hard to explore different solutions to find a better setting, since it would require experiments involving multiple runs to observe the impact on changes.

\begin{example}
Consider once more the data set of Table 1. A user may conceivably think that the $StudentScore$ scoring function that gives to $MathGrade$ and to $EngGrade$ the same weights is a good way of determining the best top $k=2$ students. But carefully looking at data, it is easy to notice that Emma may never appear in the result even if she is the only student to have both grades greater than six.
\end{example}

\section{Skylines}
Skyline queries are not based on an aggregate scoring function with weights, but instead find their roots in the concept of pareto dominance. Again, as for top-k queries, the definition of Problem 1 has to be adapted \cite{article7}.

\begin{problem}
Given N objects described by d (dimension) attributes in a data set and given some convention for assessing the attributes (e.g. usually lower values are considered better): retrieve all objects which are not dominated by any other object.
\end{problem}
As seen, the notion of pareto dominance, or simply dominance, is key to understanding skyline queries. \cite{article3, article6, article7, article9, article12, article13, article14}
\begin{definition}
Consider the tuples $t$ and $s$ belonging to a data set with schema $R$ over $d$ dimensions and consider lower values as better. Then we can say that $t$ dominates $s$ if for each attribute, $t$ is never worse (greater or equal) than $s$ and if for at least one attribute, $t$ is strictly better (less) than $s$.
\end{definition}
As a result of Definition 1, skylines can now be properly defined as well \cite{article2, article16}:

\begin{definition}
A skyline is the set of all non-dominated tuples or, alternatively and equivalently, the set of potentially optimal tuples.
\end{definition}
An example is included to show how skyline works. It will be used further to discuss some characteristics of this technique.

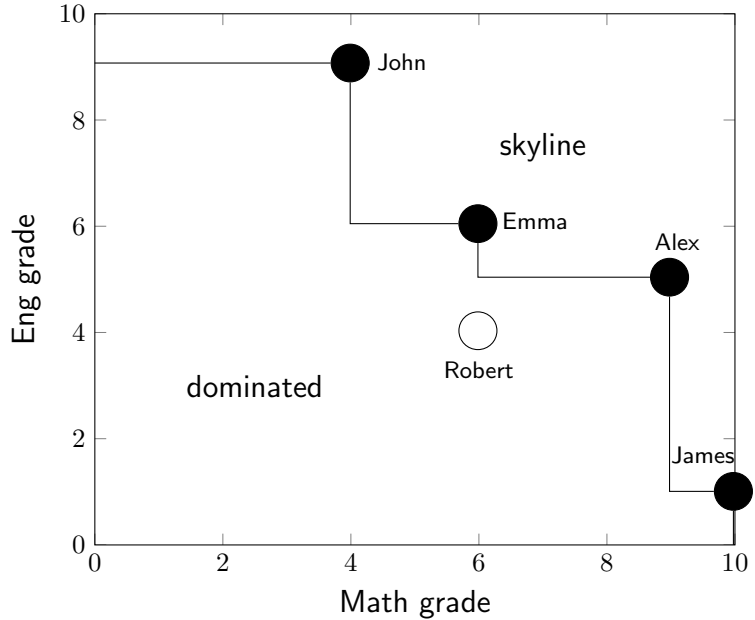
\begin{figure}
    \begin{center}
        \begin{tikzpicture}[font=\sffamily,
                myCircle/.style={circle, draw, minimum size=5mm},
                rectCircle/.style={myCircle, fill=black}
            ]
        \begin{axis}[
            title={},
            xlabel={\large Math grade},
            ylabel={\large Eng grade},
            xmin=0, xmax=10,
            ymin=0, ymax=10,
            xtick={0,2,4,6,8,10},
            ytick={0,2,4,6,8,10},
            ymajorgrids=false,
            no markers,
        ]      
            \node[] at (axis cs: 2.5,3) {\large dominated};
            \node[] at (axis cs: 7,7.5) {\large skyline};
            \node[] at (axis cs: 4.8,9.1) {\small John};
            \node[] at (axis cs: 6.88,6.1) {\small Emma};
            \node[] at (axis cs: 9.1,5.7) {\small Alex};
            \node[] at (axis cs: 9.5,1.7) {\small James};
            \node[] at (axis cs: 6,3.3) {\small Robert};
        \end{axis}
            \node [rectCircle] (t1) at (7.56,3.55) {};
            \node [rectCircle] (t2) at (8.4,0.71) {};
            \node [rectCircle] (t5) at (5.04,4.26) {};
            \node [rectCircle] (t3) at (3.36,6.39) {};
            \node [myCircle] () at (5.04,2.84) {};
            \coordinate (orig) at (0,0);
        \draw [](8.4,0) -- (t2) -- (7.56,0.71) -- (t1) -- (5.04,3.55) -- (t5) -- (3.36,4.26) -- (t3) -- (0,6.39);
        \end{tikzpicture}
        \caption{
            Skyline for tuples from Table 1. Higher values are better. The black dots are part of the skyline.
        }
    \end{center}
\end{figure}

\begin{example}
We take as landmark the data set in Table 1 and then apply the skyline operator. The results are shown graphically in Figure 1. The skyline is composed of all students that are not dominated (e.g. no other student has a greater or equal grade in both Math and Eng with at least one of the two strictly greater). Reading the plot from left to right, we can find this set of tuples in the skyline: John, Emma, Alex, James. The white dot, representing Robert, is dominated by both Emma and Alex, thus it's excluded.
\end{example}
Before moving on to the next section, it is worth noting that the concept of skyline can be further generalized by introducing the concept of k-skybands \cite{article3, article4}:

\begin{definition}
A k-skyband is the set of all tuples dominated by less than k other tuples.
\end{definition}
It's easy to note that a skyline is nothing else than a k-skyband with a particular value of $k=1$. Another interesting aspect to consider consists in the fact that the result of every top-k query is contained in the k-skyband query executed on the same data set. Many k-skyband based approaches exist for solving different problems (e.g. the one in \cite{article10}).

\subsection{Pros and cons}
Skylines completely solve the weights problem and this approach does not require them to work properly \cite{article1, article2, article3, article6}. Actually, this approach is more intuitive since it adapts itself to the multi-criteria optimization problem while ranking queries transform the problem in a single-objective optimization one \cite{article6}.

Another advantage consists in the greater informative quality of the result. Considering Example 3, it's clear that all possible interesting tuples are included in the result. Regrettably, there is another side of the coin. The size of the result may increase too much especially when dealing with many attributes $d$ and in the worst case it may reach the size of the entire data set \cite{article7, article9, article12, article14}. The average user may be overwhelmed by the task of choosing between many tuples \cite{article6, article12, article14}

Skylines are said to be more complex than ranking queries and usually a little more expensive to compute but that doesn't seem enough to overshadow the previously mentioned advantages \cite{article1,article6}. Furthermore, the authors of \cite{article14} provide an algorithm to parallelize the computation of skylines making this technique suitable also in distributed scenarios.

Another drawback of this technique consists in the fact that it lacks a feature for adjusting the relative importance of each attribute \cite{article9, article13}.

\begin{example}
Consider the results obtained in Example 3. The skyline operator gave us all the possible interesting results. We have Emma with a balance between grades, James is the best in Math, Alex is good in Math and not so bad in Eng and finally John has opposite grades to Alex. A user dealing with these tuples may have a hard time choosing the best student. In a simple data set as the one shown in Table 1, the skyline approach gave us almost the entire data set, hence it's evident that the situation could have been a lot worse in a real world scenario.
\end{example}

\section{Related work}
In the previous sections disadvantages exhibited by ranking and skyline queries have been deeply described. The scientific community has designed multiple solutions to solve those problems. Usually, each of the solutions start from one of the two classic methods and then either tries to mix them up or consider adding new strategies. 

In \cite{article11}, the authors introduced an instance optimal algorithm called FSA to compute top-k queries that do not require precise values of weights but just a rough estimate. Their work is based on the concept of F-dominance that in this document will be dealt with in greater depth in the next section where F-skylines are analyzed.

Another possibility is shown in \cite{article12}, where top-k dominating queries have been studied. This technique consists in ranking queries with a particular scoring function which, considering a tuple $t$ consists in the count of tuples dominated by $t$. An optimized algorithm has been provided too. 

A further study \cite{article4} has presented uncertain top-k queries (UTK) in which, instead of avoiding the weighted approach, the provided algorithm takes into account the weight uncertainties. The resulting query works using the concept of dominance and tries to include in the result set also interesting results within a certain range from the combination of weights.

Next, moving into more skyline based methods, there is \cite{article9} that adds attribute importance (preferences) to skylines creating the p-skyline framework. Incorporating preferences also reduces considerably the result size. Unfortunately, this approach somehow requires user preferences and limits the flexibility between dimensions.

An additional extension is the trade-off skyline paradigm \cite{article13}. As the name suggests, they aim to introduce the concept of trade-off between attributes in skylines. Despite this, the proposed algorithm is not so efficient and the trade-off preferences require to be somehow elicited.

All presented solutions solve either some or even a great part of the common problems that are usually encountered in multi-objective optimization, but they all have some drawbacks.

\section{Flexible skylines}
Every approach is imperfect, but flexible skylines appear to solve most of the issues with fewer problems compared to other methods. Their focus is to enhance basic skylines by giving control of the result size \cite{article1}. Originally, they were called restricted skylines (or r-skylines) \cite{article1, article2, article7} but newer literature refers to them as flexible skylines \cite{article3}. Their foundation consists in the decision of weakening the notion of pareto dominance \cite{article7}. This process is accomplished by imposing a set of constraints on the scoring functions \cite{article1, article2, article3}. As a result, the notion of F-dominance has been obtained:
\begin{definition}
Consider the tuples $t$ and $s$ belonging to a data set with schema $R$ over $d$ dimensions. Then we can say that $t$ $F$-dominates $s$ if for each considered scoring function, $t$ is never worse than $s$ and if for at least one scoring function, $t$ is strictly better than $s$.
\end{definition}
The concept of F-dominance helps to greatly reduce the result set with respect to the classic dominance. As a consequence of including the set of constraints, F-skylines can adapt to user preference and hence they are more flexible.

The definition of F-dominance leads to the definition of two different F-skylines operators. They are respectively called ND and PO operators.
\begin{definition}
An ND F-skyline is the set of all non F-dominated tuples.
\end{definition}

\begin{definition}
A PO F-skyline is the set of all potentially optimal tuples (i.e. top 1 according to some scoring function).
\end{definition}
It is worth noting an interesting property of the F-skylines operators \cite{article1, article3}:
\begin{equation}
PO \subseteq ND \subseteq SKY
\end{equation}
This shows clearly that the size of PO and ND result sets has to be smaller than the size of the classic skyline (equal in the worst case). Additionally, as said before, F-skylines offer the necessary flexibility that was missing in the basic skylines and do not require precise weights (only constraints). It is clear that F-skylines solve a great portion of the problems described in previous sections.
\begin{example}
Consider the data set of Table 1 again and $F$ as the set of linear scoring functions of the form $StudentScore=w_{1}*MathGrade+w_{2}*EngGrade$ such that $w_{1} > w_{2}$. The PO result set is composed by the students Alex and James. As expected, the result size is smaller than the one of the basic skyline.
\end{example}
Before moving on, there is another thing to take into account that our example highlights. F-skylines successfully manage to reduce the size of the result set, though they can't estimate the exact size of the result before computing it. This issue has been addressed in \cite{article5} where the two operators ORD and ORU have been designed.
Last but not least, the assessment of performances can be found in the next subsection.

\subsection{Algorithms}
In \cite{article3} several algorithms to compute F-skylines have been introduced. The most efficient between them are reported here along with a brief description.

Computing ND highly depends on how challenging the data set is. In case of a challenging scenario, it is advised to use the SVE1F algorithm while in the opposite case it is more suitable the SVE2 algorithm. The names of the algorithms contain information to understand their behaviour.
\begin{enumerate}
  \item The S means that the algorithm Sorts the input data set, which is always beneficial in ND computation.
  \item The VE part means that the algorithm works with the Vertex Enumeration strategy which is usually faster than applying linear programming.
  \item The digit points out the number of phases required. Number 2 means that ND is computed from the classic skyline.
  \item The F means that the algorithm interleaves dominance and F-dominance during its execution.
\end{enumerate}
The best method to compute PO is always starting from ND and then applying the PODI2 algorithm, which behaves well in all scenarios.
\begin{enumerate}
  \item The PO part means that the algorithm computes the PO operator.
  \item The D part means that F-dominance is checked through a dual PO test which usually leads to better performances than the primal test.
  \item The I means that an incremental approach is being used.
  \item The digit points out the number of phases required. Number 2 means that PO is computed from the ND F-skyline.
\end{enumerate}
In the cited article it is proven that the performances of F-skylines are usually better than normal skylines (computed with the famous SFS \cite{article3, article14}) when computing ND and in line with them for what concerns PO. This confirms the fact that F-skylines can be used in place of normal skylines without worrying. Unfortunately, they lose in the comparison of computational time with ranking queries. The increased capability of exploring interesting results comes with a price. However, as for basic skylines, F-skylines can be extended to work in a parallel or distributed environment.

\section{Conclusion}
This paper has presented a list of approaches for dealing with multi-objective optimization in data sets from classic methods to newly designed ones. This review has shown how top-k queries and skylines work along with their advantages and disadvantages, highlighting their weaknesses. Next, a list of methods aiming to overcome the previously mentioned difficulties has been presented, up to the flexible skyline paradigm, which solves most of (unfortunately not all) the problems that arise with others solutions. F-skylines have been deeply analyzed and some of the most efficient algorithms to compute them have been presented. Another interesting approach that has been mentioned but not investigated in detail is the ORD and ORU operators that managed to solve one of the few F-skylines drawbacks.

In conclusion, even if flexible skylines are not a panacea for multi-objective optimization, it sure is one of the most valid approaches and it is worth keeping an eye on their possible evolution and applicability. In the future, investigating further optimizations of this method may be interesting and the remaining drawbacks of this approach may be solved.

\bibliographystyle{plain}
\bibliography{refs.bib}
\end{document}